\documentclass[aps,prl,epsf,epsfig,twocolumn]{revtex4}
\usepackage[dvips]{graphicx}

\begin{document}

\title{Far-from-equilibrium growth of thin films in a temperature gradient} 

\author{Juli\'an Candia$^1$ and Ezequiel V. Albano$^{1,2}$}

\affiliation{1 - Instituto de F\'{\i}sica de L\'{\i}quidos y Sistemas 
Biol\'ogicos,\\ (CONICET, UNLP), 59 Nro 789, 1900 La Plata, Argentina\\
2- Departamento de F\'{\i}sica (UNLP), La Plata, Argentina}

\begin{abstract}
The irreversible growth of thin films under far-from-equilibrium conditions is 
studied in $(2+1)-$dimensional strip geometries. 
Across one of the transverse directions, a temperature gradient is applied by thermal baths at fixed temperatures 
between $T_1$ and $T_2$, where $T_1<T_c^{hom}<T_2$ and $T_c^{hom}=0.69(1)$ is the critical temperature 
of the system in contact with an homogeneous thermal bath. 
By using standard finite-size scaling methods, we characterized a continuous order-disorder phase transition 
driven by the thermal bath gradient with critical 
temperature $T_c=0.84(2)$ and critical exponents $\nu=1.53(6)$, 
$\gamma=2.54(11)$, and $\beta=0.26(8)$, 
which belong to a different universality class from that of films grown in an homogeneous bath. 
Furthermore, 
the effects of the temperature gradient are analyzed by means of a bond model that captures 
the growth dynamics. The interplay of geometry and thermal bath asymmetries leads to growth bond flux 
asymmetries and the onset of transverse ordering effects that explain qualitatively 
the shift in the critical temperature. 
\vskip 0.3 true cm
\noindent PACS numbers: 81.15.Aa, 68.35.Rh, 64.60.De, 05.50.+q
\end{abstract}

\maketitle

From an experimental and applied science perspective, 
thin films are extremely useful in a variety of areas, from the manufacture of optics (for reflective 
and anti-reflective coatings, self-cleaning glasses, etc.) to electronics (layers of insulators, 
semiconductors, and conductors from integrated circuits) and packaging (e.g. aluminium-coated PET films). 
Since the growth temperature is one of the key parameters in the formation of ordered thin films, 
several experiments have focused on the influence of a temperature gradient during film growth.
In an early experiment by Tanaka {\it et al.}~\cite{tana87}, magnetic Tb-Fe films were grown between two 
substrates with a temperature gradient, reporting the observation of perpendicular 
magnetic anisotropies and other gradient-driven structural features. More recently, 
Schwickert {\it et al.}~\cite{schw00} introduced the {\it temperature wedge method} where a calibrated 
temperature gradient of several hundred Kelvin was established across the substrate during co-deposition 
of Fe and Pt on MgO(001) and MgO(110) substrates. 
Also, Yongxiong {\it et al.}~\cite{yong05} have investigated the evolution of Fe oxide 
nanostructures on GaAs(100) by using a multi-technique experimental setup. 
In these studies, nanoscale epitaxial 
Fe films were grown, oxidized, and annealed using a gradient temperature method, which led to 
nanostripes with uniaxial magnetic anisotropy. Moreover, 
the experimental progress on this field has led to a variety of technological applications as well~\cite{kawaguchi}.  

From a theoretical standpoint, the so-called {\it gradient percolation method} was originally introduced to 
investigate the percolation transition and later applied to a variety of 
problems~\cite{gouy05}, as e.g. first- and second-order 
irreversible phase transitions in far-from-equilibrium systems~\cite{losc09}. 
In magnetic systems, damage spreading processes in a temperature gradient~\cite{bois91} 
and studies of several one-dimensional models~\cite{sait96} 
have been followed by the investigation of the kinetic Ising model in two dimensions under different kinds of dynamics~\cite{casa07}.

In the context of these recent experimental and theoretical investigations, this work focuses on 
the irreversible growth of magnetic thin films in a transverse temperature gradient by means of extensive 
Monte Carlo simulations. To our best knowledge, this is the first investigation on the growth 
of magnetic films that presents a full characterization of a gradient-driven phase transition 
and universality class.  
Magnetic films growing under far-from-equilibrium conditions 
are investigated by using the magnetic Eden model~\cite{ausl93}, an extension 
of the classical Eden model in which particles have an additional degree of freedom (the spin).  
Notice that, despite the fact that we use a magnetic terminology throughout, this work is not restricted 
to magnetic films and can be applied to other physical systems, e.g. binary alloys.
Earlier studies have shown that films growing in $(d+1)$-dimensional strip  geometries in an homogeneous thermal bath are 
noncritical for $d=1$~\cite{cand01}. 
However, for $d=2$ they undergo order-disorder phase transitions that take place at 
$T_c^{hom}=0.69(1)$. Intriguingly, the critical exponents 
for this far-from-equilibrium growth model agree within error 
bars with the exact exponents for the equilibrium Ising model in $d=2$~\cite{cand01}. 

In this work, magnetic films in $(2 + 1)-$dimensions are studied in a square-lattice 
geometry $L_x \times L_y\times L_z$, where $L_z\gg L_x=L_y\equiv L$ 
is the growth direction. The starting seed for the growing film is a plane of $L \times L$ up spins (i.e. $S=1$) placed at $z=1$ 
and film growth takes place along the positive longitudinal direction ($z\geq 2$). 
The growth process is characterized by an initial transient length $\ell_T\sim L^2$ followed by a nonequilibrium stationary state that is independent of the initial configuration \cite{cand01}. Therefore, any choice for the seed leads to the same stationary states for $z\gg\ell_T$. By disregarding the transient region, all results reported in this paper are 
obtained under stationary conditions. 
Across one of the transverse directions (the $y-$axis), a temperature gradient is applied by 
thermal baths at fixed temperatures linearly varying between $T_1$ and $T_2$. Hence, we adopt open boundary 
conditions along the $y-$direction. Across the other transverse direction (the $x-$axis), continuous boundary conditions are considered, meaning that sites at $(x=1,y,z)$ and $(x=L,y,z)$ are lattice neighbors. Notice that the 
assumption of continuous boundary conditions is the standard method used in Monte Carlo simulations 
to eliminate the impact of surface effects that would arise otherwise due to missing neighbors in the lattice 
boundaries ~\cite{bind00}.  
Unless stated otherwise, results below correspond to $T_1=0.5$ and $T_2=1.5$. We checked that 
changing the gradient has only mild finite-size effects that become negligible in the thermodynamic limit. That is, 
by applying different gradients the system's observables such as the magnetization, susceptibility and higher-order cumulants 
are shifted in the transition region by small amounts, which vanish in the $L\to\infty$ limit. More details 
with a full account of finite-size effects will be published elsewhere~\cite{cand11}.  

Films are grown by selectively adding spins ($S_{\bf{r}}=\pm 1$) to perimeter sites, 
which are defined as the nearest-neighbor (NN) empty sites of the already occupied ones. Let us recall that the substrate is a $3D$ cubic lattice and therefore each lattice site has 6 NN sites. The deposition of new spins is 
irreversible (i.e. once added, spins do not flip, detach, nor diffuse).  
Considering a ferromagnetic interaction of strength $J>0$ between NN spins, 
the energy of a given configuration of spins is given by 
$E=-\left(J/2\right)\sum_{\langle {\bf{r}},{\bf{r'}}\rangle}S_{\bf{r}}S_{\bf{r'}}$,
where the summation is taken over occupied NN sites.  
The Boltzmann constant is set equal to unity throughout, 
and both temperature and energy are measured 
in units of $J$. The probability for a perimeter site to be occupied is proportional to the Boltzmann factor
$\exp(- \Delta E/T)$, where $\Delta E$ is the change of energy involved in the addition of
the new particle and $T$ is the temperature at the perimeter site. 
At each step, all perimeter sites have to be considered and the probabilities 
of adding a new (either up or down) spin to each site must be evaluated. 
Since this procedure requires updating the probabilities at 
each time step, the algorithm is very slow compared with those used in equilibrium spin models.
Involving a significant computational effort, 
clusters having up to $10^9$ spins have been generated for lattice sizes in the range $12\leq L\leq 96$.  
As in the case of the classical Eden model, the magnetic Eden model 
leads to a compact bulk and a self-affine growth interface~\cite{ausl93}. 
Snapshots and further details on the growth behavior will be given elsewhere~\cite{cand11}. 

In order to study magnetic films growing in a gradient, the appropriate order parameter
is the mean absolute magnetization of transverse columns at constant temperature, i.e. 
$\langle|m|\rangle(y)=\left( 1/L\right)\langle |\sum_xS_{xyz}|\rangle$, 
where $\langle ...\rangle$ denotes averages along the growth direction $z$, and 
the absolute value avoids shortcomings due to finite-size effects, as in standard 
Monte Carlo simulations~\cite{bind00}.  

\begin{figure}[t!]
\includegraphics[width=9.0cm,clip=true,angle=0]{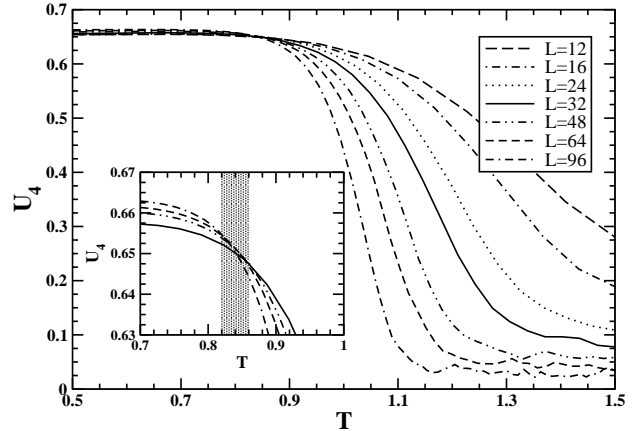}
\caption{Binder cumulant as a function of the layer temperature for  
different system sizes, as indicated. 
The inset shows the cumulant intersections for the larger systems 
($32\leq L\leq 96$), which determine $T_c=0.84(2)$.}
\label{cumulant}
\end{figure}

The Binder cumulant, $U_4(y)\equiv 1-\langle m^4\rangle/3\langle m^2\rangle^2$, is useful to 
locate the critical temperature because, in the low-temperature 
ordered region, $U_4$ tends to the value $2/3$, while in the high-temperature 
disordered region, $U_4$ tends to $0$. Thus, in the thermodynamic 
limit, $U_4$ becomes discontinuous exactly at the critical temperature \cite{bind81}. 
Figure~\ref{cumulant} shows $U_4$ as a function of the layer temperature for different system sizes. 
Notice that, by our setup assumptions, each layer at fixed $y$ is subjected to a constant 
temperature $T(y)=T_1+(T_2-T_1)(y-1)/(L-1)$ maintained by a thermal bath. We assume that each layer's temperature can be 
maintained at the same value through the film's growth process (that is, we assume that the 
deposition process does not affect the layer's temperature, which is fixed by that layer's thermal bath). 
The results from Figure~\ref{cumulant} show evidence for the existence of a continuous phase transition 
driven by the temperature gradient. 
Indeed, if a non-trivial critical temperature exists in the thermodynamic limit, 
we should expect the cumulants for different lattice sizes to intersect near the critical temperature~\cite{bind81}.  
The inset in Figure~\ref{cumulant} 
shows a detailed view of the data for the largest lattice sizes available, where 
the intersection region is also indicated. Based on this observation, we determine 
the critical temperature in the thermodynamic limit as $T_c=0.84(2)$. 

\begin{figure}[t!]
\includegraphics[width=8.8cm,clip=true,angle=0]{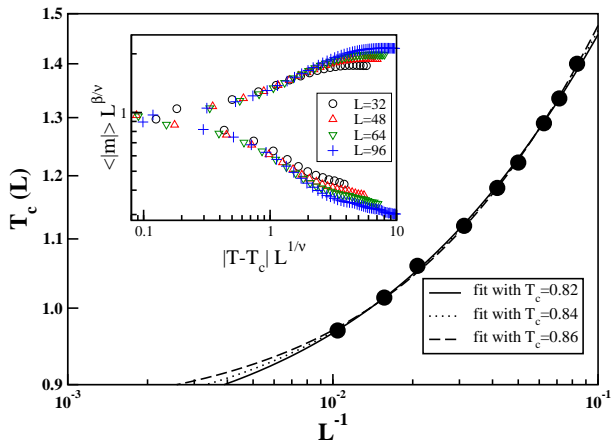}
\caption{(Color online) Log-log plot of the finite-size pseudo-critical temperature $T_c(L)$ as a function of the 
inverse of the system linear size, $L^{-1}$. Finite-size scaling fits to the data obtained by 
using $T_c=0.82, 0.84,$ 
and $0.86$ lead to the critical exponent $\nu=1.53(6)$. Inset: plot of 
$\langle |m|\rangle\times L^{\beta/\nu}$ vs 
$|T-T_c|\times L^{1/\nu}$ (with $\beta=0.26$) 
showing a data collapse for different system sizes in the range $32\leq L\leq 96$.}
\label{nu}
\end{figure}

In the following, we apply standard finite-size scaling techniques~\cite{bind00,priv90} 
to determine the critical exponents that 
characterize the system's critical behavior and universality class. 
Figure~\ref{nu} shows a log-log plot of the   
finite-size pseudo-critical temperature $T_c(L)$ as a function of the 
inverse of the system linear size, where $T_c(L)$ is defined as the temperature corresponding 
to $\langle |m|\rangle=0.5$. The finite-size scaling theory predicts that $|T_c-T_c(L)|\propto L^{-1/\nu}$,  
where $\nu$ is the exponent that characterizes the divergence of the correlation length 
at criticality. 
Least-squares fits to the data obtained by using $T_c=0.82, 0.84,$ and $0.86$ are also shown. 
We observe that the finite-size scaling relationship fits the data very well within the range of values for 
the critical temperature that was derived from the intersection of Binder's cumulants. 
From these fits, we obtain the critical exponent $\nu=1.53(6)$, 
where the error bars reflect the error derived from the evaluation of $T_c$ as well as the statistical error. 
On the other hand, the finite-size
scaling theory predicts that plots of $\langle |m|\rangle L^{\beta/\nu}$ vs 
$|T-T_c| L^{1/\nu}$ for different lattice sizes should collapse near the critical region. 
The inset to Figure~\ref{nu} shows the data collapse obtained by 
using $\beta=0.26$ (that is determined from the hyperscaling relation, as explained below)
with two separate branches corresponding to the low- and high-temperature regions.  

\begin{figure}[t!]
\includegraphics[width=9.0cm,clip=true,angle=0]{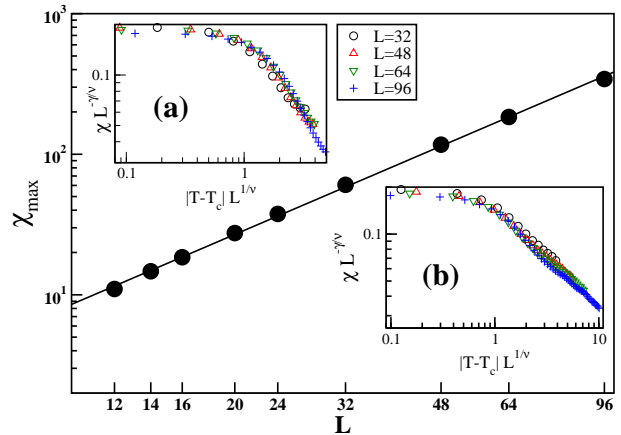}
\caption{(Color online) Log-log plot of the susceptibility maxima as a function of the system linear size, where 
the solid line is a finite-size scaling fit that yields $\gamma/\nu=1.66(3)$. The insets display  
plots of $\chi L^{-\gamma/\nu}$ vs $|T-T_c| L^{1/\nu}$ showing separately the data collapse for 
(a) the low-temperature branch and (b) the high-temperature branch.}
\label{susceptibility}
\end{figure}

Let us now calculate the critical exponent $\gamma$, which describes the divergence of the susceptibility 
at the critical point. The susceptibility is defined as 
$\chi = \left(L^2/T\right)\times\left(\langle m^2\rangle-\langle|m|\rangle^2\right)$. 
From the finite-size scaling theory~\cite{priv90}, 
the exponent ratio $\gamma/\nu$ is related to the peak of the susceptibility measured in finite samples 
of size $L$ by $\chi_{max}\propto L^{\gamma/\nu}$. Figure~\ref{susceptibility} shows 
the maxima of $\chi$ vs $L$, where the solid line is a fit to the data 
that yields $\gamma/\nu=1.66(3)$, where the error bars reflect the statistical error from the fit. By 
using this ratio and the value already obtained for $\nu$, we determine $\gamma=2.54(11)$. 
The insets to Figure~\ref{susceptibility} display plots of 
$\chi L^{-\gamma/\nu}$ vs $|T-T_c| L^{1/\nu}$ 
for different lattice sizes in the range $32\leq L\leq 96$. Using the critical temperature as determined by the 
susceptibility peaks, the data collapse is shown separately for (a) the low-temperature branch and (b) the high-temperature branch. In the former case, data from low-temperature layers near $T_1=0.5$ depart from the collapse 
and have been removed. However, the collapse near the critical region is remarkable and agrees very well with the expectations from the finite-size scaling theory. 

By replacing the exponents $\nu$ and $\gamma$ in the hyperscaling relation $d\nu-2\beta-\gamma=0$ with $d=2$, we determine the exponent $\beta=0.26(8)$. Indeed, we anticipated this value of $\beta$ when we considered the 
data collapse of the scaled magnetization (see the inset to Figure~\ref{nu} above). The excellent data collapse 
near the critical region confirms the consistency and robustness of the obtained results. 
Based on these findings, 
we conjecture that magnetic Eden films grown in a temperature gradient belong to a new universality class 
characterized by critical exponents $\nu=3/2, \gamma=5/2,$ and $\beta=1/4$.   
In contrast, previous studies found that the critical exponents for magnetic Eden films grown in an homogeneous bath agree within error 
bars with the exact exponents for the Ising model in $d=2$~\cite{cand01}, namely $\nu=1, \gamma=7/4$, and $\beta=1/8$. 

\begin{figure}[t!]
\includegraphics[width=9.0cm,clip=true,angle=0]{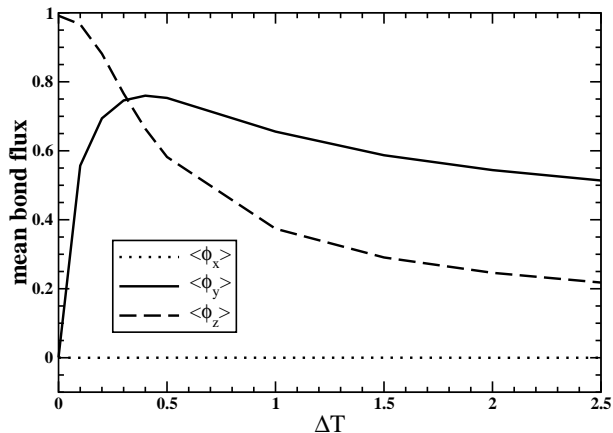}
\caption{Mean growth bond flux components as a function of the gradient span 
$\Delta T$ for $L=32$ and $T_1=0.5$. Asymmetries due to the temperature 
gradient and the substrate geometry are responsible for net bond fluxes along 
the $y$ and $z$ directions. Along the tranverse direction $x$ the system is fully symmetric, 
so no net bond fluxes are observed.}
\label{meanbondfluxes}
\end{figure}

Let us now explore the growth dynamics by means of a bond representation. For this purpose,
to each pair of neighboring sites
we assign a directed bond that points from the earlier occupied site to the later occupied one. 
The components of the bond flux field $\vec{\phi}$ at a site $\vec{s}=x\breve{\i}+y\breve{\j}+z\breve{k}$ 
are defined as $\phi_x(\vec{s})=b[\vec{s},\vec{s}+\breve{\i}]$, 
$\phi_y(\vec{s})=b[\vec{s},\vec{s}+\breve{\j}]$, and 
$\phi_z(\vec{s})=b[\vec{s},\vec{s}+\breve{k}]$, 
where $b[\vec{s_1},\vec{s_2}]=+1$ $(-1)$ if the bond points from $\vec{s_1}$ to $\vec{s_2}$ 
$(\vec{s_2}$ to $\vec{s_1})$.  
Figure~\ref{meanbondfluxes} shows the components of the 
mean bond flux as a function of the gradient span $\Delta T\equiv T_2-T_1$ for $L=32$ and $T_1=0.5$.   
As expected from the symmetry along $x$, one has that $\langle\phi_x\rangle=0$ regardless of $\Delta T$.   
For $\Delta T=0$, the system 
is also symmetric along $y$, so no net bond flux is observed. When a gradient is applied, however, 
this symmetry is broken. Since the growth probabilities depend on the Boltzmann factor 
$\exp(- \Delta E/T(y))$, where $T(y)$ is the layer's temperature, the thermal asymmetries introduced 
by the gradient favor spin deposition on the colder layers. This phenomenon is captured by the 
observed net bond flux $\langle\phi_y\rangle > 0$. Indeed, as shown in Figure~\ref{meanbondfluxes}, 
the thermal asymmetries cause $\langle\phi_y\rangle$ to grow steeply up to 
$\langle\phi_y\rangle_{max}\approx 0.75$ followed by a moderate decrease for larger gradients, 
which is due to the onset of bulk disorder within the hotter layers. 
Since the net transverse growth bond flux is directed from the ordered (cold) layers 
towards the disordered (hot) ones, 
this gradient-induced transverse ordering mechanism causes the system's critical temperature to increase 
from $T_c^{hom}=0.69(1)$ to $T_c=0.84(2)$. 
On the other hand, for $\Delta T=0$, $\langle\phi_z\rangle=1$ 
due to the longitudinal asymmetries in the substrate (i.e., the semi-infinite strip geometry constrains 
the system to grow along the $z>0$ direction). 
However, when the transverse gradient is applied, two effects contribute to decrease 
$\langle\phi_z\rangle$: $(i)$ the onset of the transverse bond flux, which creates transverse domains in the 
active perimeter and causes some of the added spins to grow backwards; $(ii)$ the bulk 
disorder induced in the hotter layers 
(which also causes $\langle\phi_y\rangle$ to decrease, as discussed above).  

As a summary, we studied the growth of thin films under far-from-equilibrium conditions 
in $(2+1)$-dimensional strip geometries, where a temperature gradient is applied across 
one of the transverse directions.   
In order to investigate the system's critical behavior, 
we applied the finite-size scaling theory. 
The critical temperature is $T_c=0.84(2)$ and the critical exponents are $\nu=1.53(6)$,  
$\gamma=2.54(11)$, and $\beta=0.26(8)$. 
Based on these findings, we conjecture that the exact exponents are $\nu=3/2$, $\gamma=5/2$, and $\beta=1/4$, 
which satisfy the hyperscaling relationship and point out the existence of a new universality class 
for film growth in a thermal gradient. 
Moreover, we investigated the system's growth dynamics by means of a bond model. We found that 
the interplay of geometry and thermal bath asymmetries leads to growth bond flux 
asymmetries and the onset of transverse ordering effects that explain qualitatively 
the shift observed in the critical temperature.

We believe that this work provides 
the first full characterization of a gradient-driven phase transition of growing
films under far-from-equilibrium conditions.
Given the great experimental and theoretical interest in thin films growing in 
temperature gradients, 
as well as the wide variety of technological applications that benefit from these efforts, 
we hope that this paper will contribute to the progress of this research field and stimulate further work.  

{\bf  ACKNOWLEDGMENTS}. This work was financially supported by CONICET, UNLP, and ANPCyT (Argentina).

\end{document}